\documentclass[a4paper]{article}

\usepackage{INTERSPEECH2020}

\title{Efficient Low-Latency Speech Enhancement with Mobile Audio Streaming Networks}
\name{Michał Romaniuk, Piotr Masztalski, Karol Piaskowski, Mateusz Matuszewski}
\address{Samsung R\&D Institute Poland}
\email{\{m.romaniuk2, p.masztalski, k.piaskowski, m.matuszews2\}@samsung.com}

\begin{document}

\maketitle
\begin{abstract}
We propose Mobile Audio Streaming Networks (MASnet) for efficient low-latency speech enhancement, which is particularly suitable for mobile devices and other applications where computational capacity is a limitation. MASnet processes linear-scale spectrograms, transforming successive noisy frames into complex-valued ratio masks which are then applied to the respective noisy frames. MASnet can operate in a low-latency incremental inference mode which matches the complexity of layer-by-layer batch mode. Compared to a similar fully-convolutional architecture, MASnet incorporates depthwise and pointwise convolutions for a large reduction in fused multiply-accumulate operations per second (FMA/s), at the cost of some reduction in SNR.
\end{abstract}

\section{Introduction}

Speech enhancement systems are an important component of modern communication tools such as mobile telephony, VoIP and smart earphones (hearables). They can also be used as a preprocessing step in speech recognition systems \cite{donahue2018exploring}. With recent advancements in deep learning, there has been renewed interest in the speech enhancement problem, bringing quality and intelligibility improvements over prior methods \cite{wang2018supervised}. In this work we introduce Mobile Audio Streaming Networks (MASnet), an efficient convolutional neural network architecture for speech enhancement that can be implemented in a low-latency inference mode.

Our contributions are as follows:
\begin{itemize}
\item We propose MASnet, an efficient low-latency convolutional architecture for spectrogram-based audio processing;
\item We evaluate MASnet in several size variations on the standard Valentini-Botinhao noisy speech database;
\item We show that further performance gains can be achieved by adding residual connections.
\end{itemize}

\section{Background}

The problem of speech enhancement can be approached algorithmically in several ways. The first imporant choice is that of signal representation. Some algorithms work directly with waveforms \cite{pascual2017segan, germain2018speech, luo2018tasnet, rethage2018wavenet, stoller2018wave}, while others are based on spectrogram representations computed with the short-time Fourier transform (STFT) \cite{yao2019coarse, choi2019phase}. Some other methods work with Mel-scale spectrograms \cite{donahue2018exploring}. For processing algorithms based on deep learning, the spectrogram representation has the advantage of extracting meaningful features and effectively increasing the temporal receptive fields of neural network cells. Due to the efficiency of the FFT algorithm, this is achieved with only a small cost. 


In the context of STFT-based methods, a direct approach consists of noisy spectrogram analysis followed by re-synthesis of denoised spectrograms \cite{wang2014training}. In practice, a more common method is noisy spectrogram analysis followed by synthesis of a \emph{mask} which is then multiplied with the noisy spectrogram to produce the denoised version. This approach has the advantage that the masks are thought to be easier to estimate than detailed spectrograms \cite{michelsanti2019training} and this approach has been found to work better experimentally with multilayer perceptrons \cite{wang2014training}.

The spectrogram masking approach has several further variations: the masks can be binary-valued (0 or 1), real-valued or complex-valued \cite{wang2014training, michelsanti2019training, williamson2015complex}. With binary-valued and real-valued masks it is common to reconstruct waveforms from denoised STFT magnitude and (noisy) phase of the input, but it is also possible to use the Griffin-Lim phase reconstruction algorithm \cite{griffin1984signal} or train a phase model in addition to the amplitude model \cite{afouras2018conversation}. Complex-valued masks have the advantage of being able to adjust STFT phase in conjunction with magnitude. The method proposed in this paper is based on the complex ratio masking approach.

\subsection{Low-latency speech enhancement}

Spectrogram-based speech enhancement methods often treat STFT representations as images \cite{choi2019phase, yao2019coarse}. This is fine at training-time, but at test-time and especially in low-latency applications such as telephony it is important to be able to output processed audio immediately, without having to record several seconds that could be transformed into a spectrogram.

Some speech enhancement methods achieve low latency with causal recurrent neural net (RNN) models \cite{luo2018tasnet}. Low-latency convolutional models such as Wavenet \cite{oord2016wavenet} and TCN \cite{bai2018empirical} have also been used for speech enhancement \cite{rethage2018wavenet, pandey2019tcnn}. However, these models have to learn their own feature representations and compute them at inference time, which tends to require more processing power than STFT. Causal convolutional processing of STFT spectrograms was studied by Wilson \emph{et al.} \cite{wilson2018exploring} and this idea was also combined with complex ratio masking as part of an audio-visual processing pipeline \cite{ephrat2018looking}. Their network architecture is a starting point for our work which aims to reduce its computational complexity.

Tan and Wang \cite{tan2018convolutional} proposed a causal encoder-core-decoder type network where the core is an LSTM stack. Their network operates on STFT magnitude spectra and uses striding to increase receptive field size in the frequency dimension while the LSTM provides temporal memory. In contrast, our model is based on complex-valued STFT spectra, enabling it take advantage of phase information and adjust the phase of the output spectra. Pandey and Wang \cite{pandey2019tcnn} have developed an encoder-core-decoder type architecture where the core is a sequence of temporal convolutional blocks and additional skip connections are added between the encoder and the decoder. They have also used combined depthwise and pointwise convolutions in their core blocks. However, their model operates directly on frames extracted from the waveform, rather than on spectrograms.

\subsection{Efficient convolutional architectures}

Convolutional neural networks often require high-performance hardware for fast inference, so finding network architectures that come close to state-of-the-art prediction accuracy while taking less processing power is an important research topic \cite{jin2014flattened, iandola2016squeezenet, wang2017factorized, howard2017mobilenets, sandler2018mobilenetv2}. Flattened ConvNets \cite{jin2014flattened} are based on factored convolutions. Factorized ConvNets \cite{wang2017factorized} combine intra-channel convolutions with topological subdivisioning. Squeezenets \cite{iandola2016squeezenet} are built from Fire modules, each consisting of a 1x1 convolution that squeezes the number of channels and a bank of 1x1 and 3x3 convolutions that expand it again. MobileNets \cite{howard2017mobilenets} replace standard convolutions with pairs of depthwise and pointwise (1x1) convolutions. Our proposed MASnet architecture borrows from MobileNets~\cite{howard2017mobilenets} by replacing standard convolutions with pairs of depthwise and pointwise convolutions.

\section{Model details}

\subsection{Feature representation}

\begin{figure}[tb]
\vskip 0.2in
\begin{center}
\centerline{\includegraphics[trim={1.3in, 7.8in, 1.1in, 1.1in}, clip, width=\columnwidth]{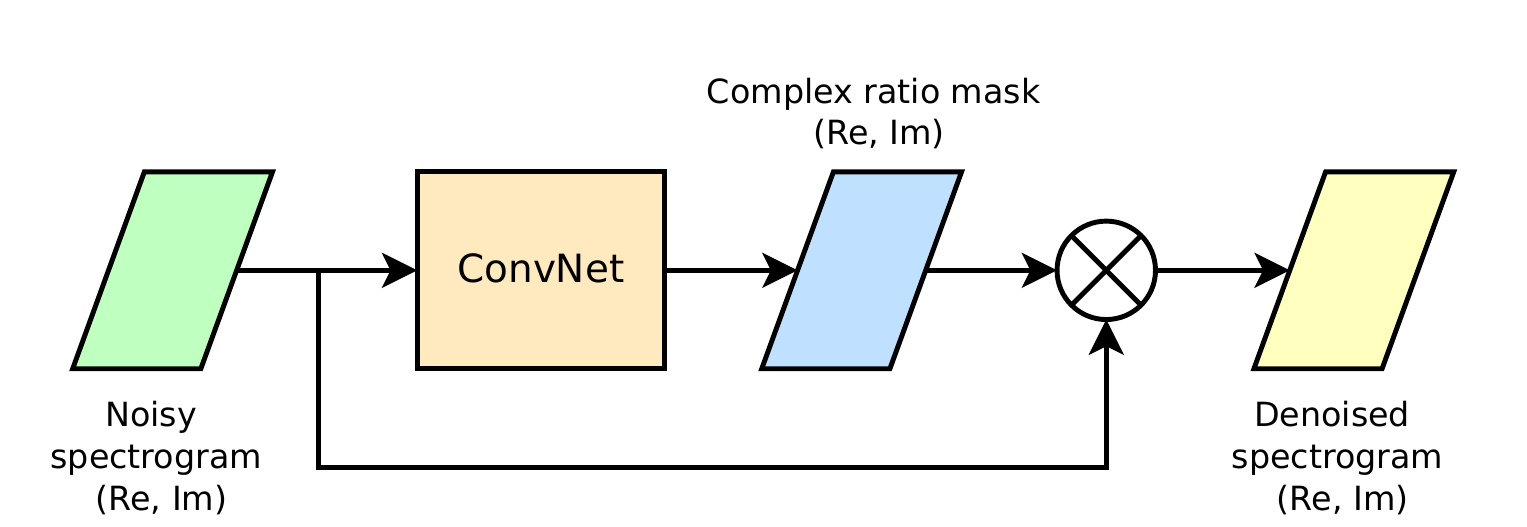}}
\caption{Model overview}
\label{model-overview}
\end{center}
\vskip -0.2in
\end{figure}

The models considered in this paper are all designed to process spectrograms represented as two-channel (real and imaginary) images and output two-channel images which are interpreted as real and imaginary parts of complex-valued ratio masks an then multiplied with the input spectrograms to produce denoised output spectrograms. The output spectrograms are then turned into output waveforms with inverse STFT. The spectrogram representations are computed with the (one-sided) STFT, using a Hann window with size 256 and stride 128. This gives 129 frequency bins. For input sampled at 16kHz, the frame rate is 125 per second.

\subsection{LLASnet}

We adopt the causal convolutional stack from \cite{wilson2018exploring} (without the LSTM and FC layers) and refer to it as Low-Latency Audio Streaming network (LLASnet). Detailed layer configurations of the LLASnet layers in two size versions: LLASnet-8 and LLASnet-15, are shown in table \ref{table_llasnet15}. All convolutions are computed with stride 1 and dilations are used to increase the receptive fields. Zero-padding is added in the convolutions as necessary to ensure that feature maps all have the same dimensions as the input frequency grid. At training time subsequent layers are computed in the conventional layer-wise way and causality is enforced by adding all the necessary zero-padding at the start of the time axis, with no padding at the end. The frequency axis is zero-padded equally on both ends. Hidden layers all use batchnorm \cite{ioffe2015batch} followed by ReLU activations. The output layers use a linear activation without batchnorm. These models are the baseline for our further investigations. We show that LLASnet can approach the state-of-the-art CRMRN U-net \cite{choi2019phase} in terms of speech quality metrics while achieving low latency, but at the cost of more computation.


\begin{table}[tb]
\caption{LLASnet-15 layer configuration (kernel shape and dilation in [time $\times$ frequency] format). Layers marked with an asterisk form the LLASnet-8 version.}
\label{table_llasnet15}
\begin{center}
\begin{small}
\begin{tabular}{lcccr}
\toprule
\textbf{Type} & \textbf{Channels} & \textbf{Kernel} & \textbf{Dilation} \\
\midrule
Conv2D* & 32 & 1x7 & 1x1 \\
Conv2D* & 32 & 7x1 & 1x1 \\
Conv2D* & 32 & 5x5 & 1x1 \\
Conv2D* & 32 & 5x5 & 2x1 \\
Conv2D* & 32 & 5x5 & 4x1 \\
Conv2D* & 32 & 5x5 & 8x1 \\
Conv2D* & 32 & 5x5 & 16x1 \\
Conv2D & 32 & 5x5 & 32x1 \\
Conv2D & 32 & 5x5 & 1x1 \\
Conv2D & 32 & 5x5 & 2x2 \\
Conv2D & 32 & 5x5 & 4x4 \\
Conv2D & 32 & 5x5 & 8x8 \\
Conv2D & 32 & 5x5 & 16x16 \\
Conv2D & 32 & 5x5 & 32x32 \\
Conv2D* &  2 & 1x1 & 1x1 \\
\bottomrule
\end{tabular}
\end{small}
\end{center}
\end{table}

\subsection{MASnet}

LLASnet can be implemented in a low-latency inference mode but it requires too much processing power to be practical in mobile or low-power applications. Our proposed model, the Mobile Audio Streaming Network (MASnet), is a further development of LLASNet, reducing the computation required. We incorporate the ideas from MobileNets \cite{howard2017mobilenets} to reduce the number of fused multiply-accumulates per second (FMA/s) by a factor of roughly 10.

The main change is to replace each convolution (including its batchnorm and ReLU) with a MAS block\footnote{This block is called a \emph{depthwise separable convolution} in the original MobileNets paper.}, consisting of a depthwise convolution followed by a pointwise (1x1) convolution, each of them followed by batchnorm and ReLU activations. Before the first layer of the network, an additional 1x1 convolution is added to expand input channel count from 2 to 32 so that the first depthwise convolution can run on 32-channel input. Applying these changes to LLASnet-8 and LLASnet-15 produces MASnet-9 and MASnet-16, respectively. These new networks are summarized in table~\ref{table_masnet16}. We also explore deeper MASnet versions by appending further layers to MASnet-16. The configuration of these extra layers is a repetition of the last 6 layers of MASnet-16 (marked with asterisks in table~\ref{table_masnet16}). By repeating this sequence 1, 2 or 3 times, we get MASnet-22, MASnet-28 and MASnet-34 respectively.


We also explore deeper MASnet versions by appending further layers to MASnet-16. The configuration of these extra layers is a repetition of the last 6 layers of MASnet-16 (marked with asterisks in table~\ref{table_masnet16}). By repeating this sequence 1, 2 or 3 times, we get MASnet-22, MASnet-28 and MASnet-34 respectively.


\begin{table}[tb]
\caption{MASnet-16 layer configuration (kernel shape and dilation in [time $\times$ frequency] format). Layers marked with $\dagger$ form the MASnet-9 version. By repeating the sequence of layers marked with asterisks we get MASnet-22, MASnet-28, MASnet-34 etc.}
\label{table_masnet16}
\vskip 0.15in
\begin{center}
\begin{small}
\begin{tabular}{lcccr}
\toprule
\textbf{Type} & \textbf{Channels} & \textbf{Kernel} & \textbf{Dilation} \\
\midrule
Conv2D$^\dagger$     & 32 & 1x1 & 1x1 \\
MAS block$^\dagger$  & 32 & 1x7 & 1x1 \\
MAS block$^\dagger$  & 32 & 7x1 & 1x1 \\
MAS block$^\dagger$  & 32 & 5x5 & 1x1 \\
MAS block$^\dagger$  & 32 & 5x5 & 2x1 \\
MAS block$^\dagger$  & 32 & 5x5 & 4x1 \\
MAS block$^\dagger$  & 32 & 5x5 & 8x1 \\
MAS block$^\dagger$  & 32 & 5x5 & 16x1 \\
MAS block  & 32 & 5x5 & 32x1 \\
MAS block* & 32 & 5x5 & 1x1 \\
MAS block* & 32 & 5x5 & 2x2 \\
MAS block* & 32 & 5x5 & 4x4 \\
MAS block* & 32 & 5x5 & 8x8 \\
MAS block* & 32 & 5x5 & 16x16 \\
MAS block* & 32 & 5x5 & 32x32 \\
Conv2D$^\dagger$    &  2 & 1x1 & 1x1 \\
\bottomrule
\end{tabular}
\end{small}
\end{center}
\vskip -0.1in
\end{table}

\subsection{Incremental low-latency inference}
\begin{figure}[tb]
\begin{center}
\centerline{\includegraphics[trim={0.9in, 3.4in, 3.7in, 1.1in}, clip, scale=0.75]{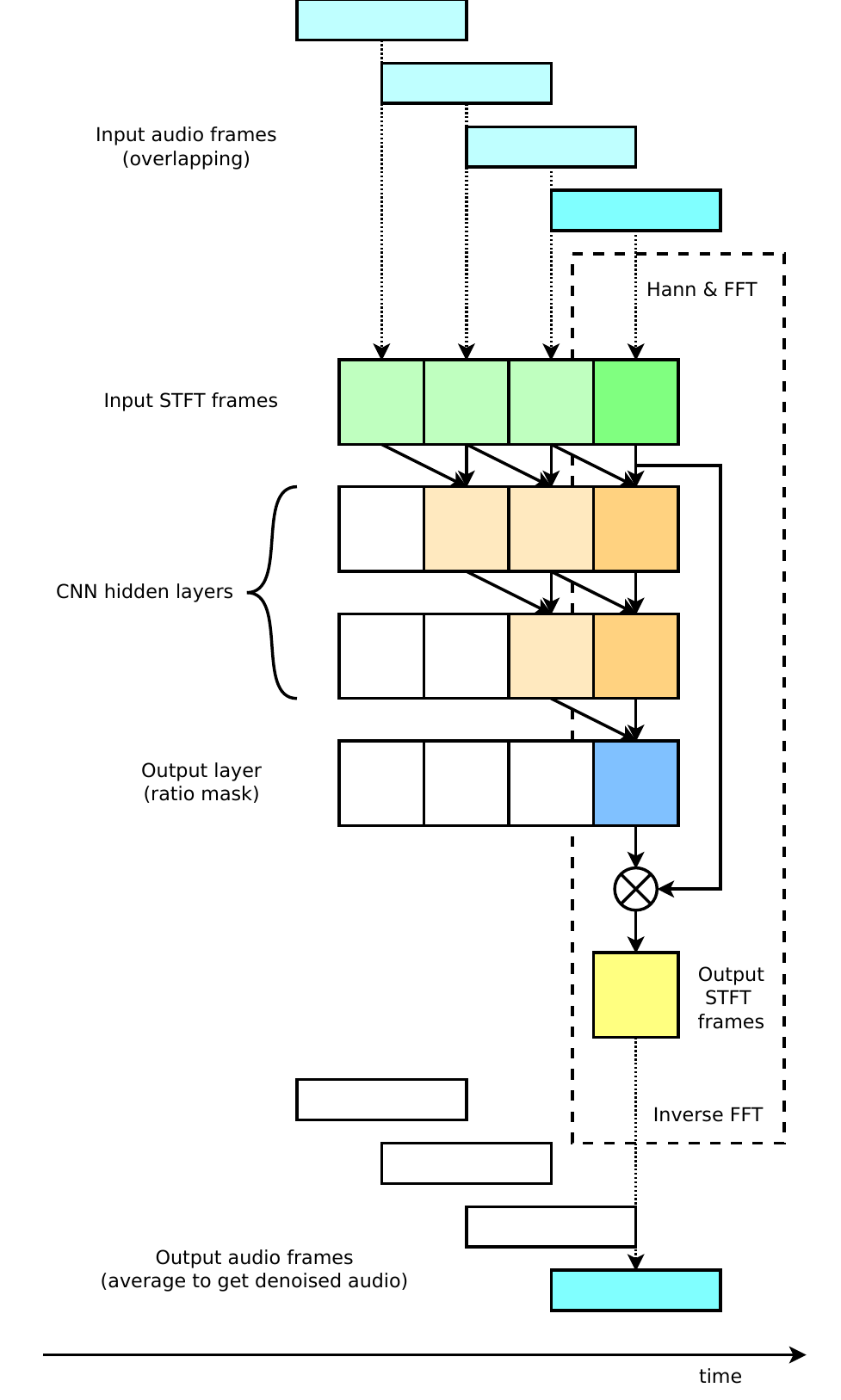}}
\caption{Low-latency inference. The dashed box contains the computations that have to be done for each new audio frame}
\label{low-latency-inference}
\end{center}
\vskip -0.2in
\end{figure}

Both LLASnet and MASnet can be implemented in an incremental low-latency inference mode, accepting one spectrogram frame at a  time and returning output immediately. This process is shown in \emph{fig.}~\ref{low-latency-inference}. Each time a new spectrogram frame arrives, we compute the corresponding frame of successive feature maps, using the cached frames from the most recent steps to reduce computational load. Due to having the same feature map dimensions for all layers, this means that we end up doing only as much computation as in layer-by-layer inference mode. While incremental inference is not very common for image processing, it has been studied before in the context of sequence models \cite{bai2018empirical, waibel1989phoneme}. 


\subsection{Training}
The network is trained by backpropagation, using the Adam optimizer \cite{kingma2014adam}. As the loss function, we use the Mean Squared Error (MSE) between the denoised spectrogram (after masking) and the clean spectrogram:

\begin{equation}
L(\hat{x}, x) = \frac{1}{TF} \sum_{t, f}^{}\left(Re\left(\hat{x} - x\right)^2 + Im\left(\hat{x} - x\right)^2\right)\label{eqLossFn}
\end{equation}

where $x$ is the clean reference spectrogram and $\hat{x}$ is the denoised estimate (noisy input multiplied by the predicted mask). The summation variables $t$ and $f$ are time and frequency respectively, while $T$ is the number of time steps (frames) and $F$ is the number of frequency bins.

\section{Experiments}

To evaluate MASnet, we use a standard dataset \cite{valentini2016investigating, valentini2016speech} and compare with LLASnet-8, LLASnet-15 and the CRMRN U-net from \cite{choi2019phase, yao2019coarse}. All of our experiments are implemented with PyTorch \cite{paszke2015pytorch}.

\subsection{Training on the Valentini-Botinhao database}
We evaluate the proposed network designs on a noisy speech database published by Valentini-Botinhao \cite{valentini2016investigating, valentini2016speech}. This dataset consists of recorded spoken phrases in two training sets: 28 speakers from England and 56 from other regions. The recordings are sampled at 48 kHz and there is an even balance between male and female speakers. Each of the clean recordings has a time-aligned noisy version with noise mixed in at SNR of 15~dB, 10~dB, 5~dB or 0~dB. The test set is based on two other speakers from England (one male and one female) and has other types of noise mixed in, with higher SNRs: 17.5~dB, 12.5~dB, 7.5~dB or 2.5~dB.

We combined the 28-speaker and 56-speker training sets into one large training set. All training and test recordings were resampled to 16kHz. To make it possible to train in batched mode, we restricted utterance length to $3 \times 16384$ samples (slightly over 3 seconds), by either truncating the recordings at the end or zero-padding them equally at the beginning and end.

In order to select a well-performing training checkpoint for evaluation, we created a validation set from the first male and first female speakers in the training set (p226 and p228). These recordings were only used for validation and not for training.

We trained all of the models for 200 epochs with batch size 16. We used the Adam optimizer \cite{kingma2014adam} with learning rate $10^{-4}$, $\beta_1 = 0.9$ and $\beta_2 = 0.999$. Then for each model we selected the best checkpoint with respect to the validation set and we report its performance on the test set.

We evaluate our models using the following metrics: Signal-to-Noise Ratio (SNR) computed on whole waveforms without segmentation\footnote{\emph{N.b.} this is different from segmental SNR (SSNR) which is also commonly used in speech enhancement}, Short-Time Objective Intelligibility (STOI) \cite{taal2011algorithm} and Perceptual Speech Quality (PESQ) \cite{rec2005p}. Table~\ref{table-llasnet-masnet} shows aggregate results.

CRMRN U-net has the most favorable performance figures and reasonable compute requirements, however it is not straightforward to adapt it for low-latency inference due to non-causal and temporally strided convolutions. In terms of SNR, LLASnet-15 comes within 1.12 dB of CRMRN U-net but its FMA/s requirement is too large for low-resource applications. LLASnet-8 requires less than half the compute of LLASNnet-15, but at the cost of a further decline in SNR of 0.9 dB. Meanwhile, MASnet-16 requires less than a tenth of the compute of LLASnet-15, with an SNR reduction of 1.41 dB. MASnet-22 requires slightly over 11\% of the FMA/s of LLASnet-15 at a cost of 1.08 dB reduction in SNR. Deeper MASnet versions seem to do worse than MASnet-22 in terms of SNR.

\begin{table}[tb]
\caption{(Table 5) Model performance comparison on Valentini-Botinhao database. SNR is in dB. All models were trained for up to 200 epochs. (Checkpoint was selected by best result on validation data.)}
\label{table-llasnet-masnet}
\vskip 0.15in
\begin{center}
\begin{small}
\begin{tabular}{lcccr}
\toprule
\textbf{Model} & \textbf{FMA/s} & \textbf{SNR} & \textbf{STOI} & \textbf{PESQ} \\
\midrule
(Noisy data)              &   --- &  8.45 & 0.9210 & 1.972  \\
\midrule
LLASnet-8                 & 2240M & 16.36 & 0.9216 & 2.383 \\
LLASnet-15                & 5199M & 17.26 & 0.9298 & 2.364 \\
\midrule
MASnet-9                  &  224M & 15.40 & 0.9155 & 2.193  \\
MASnet-16                 &  404M & 15.85 & 0.9192 & 2.213 \\
MASnet-22                 &  584M & 16.18 & 0.9219 & 2.113 \\
MASnet-28                 &  765M & 15.70 & 0.9207 & 2.035 \\
MASnet-34                 &  945M & 15.80 & 0.9192 & 2.124 \\
\midrule
CRMRN U-net               &  403M & 18.38 & 0.9333 & 2.559 \\
\bottomrule
\end{tabular}
\end{small}
\end{center}
\vskip -0.1in
\end{table}

\subsection{Adding Residual Connections}

Deeper versions of our model (MASnet-28 and MASnet-34) have worse results than MASNet-22, so we decided to see if adding residual connections \cite{he2016deep, he2016identity} would improve these deeper networks. We added an identity bypass connection to each MASblock, which resulted in a family of residual networks (MASnet-R) with a negligible increase in FMA/s over their sequential versions. The results for MASnet-R are shown in table~\ref{table-masnet-res}.

\begin{table}[tb]
\caption{(Table 6) Residual MASnet (MASnet-R) comparison on Valentini-Botinhao database. SNR is in decibels. All models were trained for up to 200 epochs. (Checkpoint was selected by best result on validation data.)}
\label{table-masnet-res}
\begin{center}
\begin{small}
\begin{tabular}{lcccr}
\toprule
\textbf{Model} & \textbf{FMA/s} & \textbf{SNR} & \textbf{STOI} & \textbf{PESQ} \\
\midrule
MASnet-R-9     &  $\approx$224M & 15.62 & 0.9185 & 2.306 \\
MASnet-R-16    &  $\approx$404M & 16.20 & 0.9184 & 2.257 \\
MASnet-R-22    &  $\approx$584M & 16.21 & 0.9246 & 2.273 \\
MASnet-R-28    &  $\approx$765M & 15.83 & 0.9224 & 2.331 \\
MASnet-R-34    &  $\approx$945M & 15.32 & 0.9192 & 2.378 \\
\bottomrule
\end{tabular}
\end{small}
\end{center}
\end{table}

Smaller MASnet-R versions (up to MASnet-R-22) have slightly higher SNR figures than their respective sequential ancestors, but the reverse is the case for MASnet-R-34. This result is somewhat surprising, since we expected residual connections to be most beneficial for larger networks. However, PESQ improves for all size versions. As for SNR and STOI, MASnet-R-22 is the best model of the MASnet family, with SNR of 16.21 dB 1.05 dB less than LLASnet-15), STOI of 0.9246 (0.0052 less than LLASnet-15) and PESQ of 2.273 (0.091 less than LLASnet-15). At the same time, MASnet-R-22 requires slightly over 11\% of the FMA operations of LLASnet-15.

\section{Conclusions}
We propose MASnet for speech enhancement applications where low latency and limited compute are important requirements. Our models achieve a large reduction in FMA/s compared to a similar fully-convolutional architecture, at the cost of some reduction in SNR. This is achieved by replacing standard convolutions with pairs of depthwise and pointwise convolutions, inspired by Mobilenets \cite{howard2017mobilenets}. It may be possible to make further performance gains by adjusing the shapes of the depthwise kernels or other architecture refinements that were proposed in the literature \cite{szegedy2015going, szegedy2016rethinking, szegedy2017inception, sandler2018mobilenetv2}, or replacing dilations in the frequency dimension with striding \cite{tan2018convolutional}. Another direction for improvement would be to add and adversarial loss \cite{goodfellow2014generative, isola2017image} in the training process, with the goal of improving the quality of denoised speech. Some work was done in this direction by other authors using diffeerent network architectures, working either with raw waveforms \cite{pascual2017segan} or with spectrograms \cite{fu2019metricgan}. However, low-latency inference was not their focus.

\bibliographystyle{IEEEtran}

\bibliography{mybib}

\begin{thebibliography}{10}
\providecommand{\url}[1]{#1}
\csname url@samestyle\endcsname
\providecommand{\newblock}{\relax}
\providecommand{\bibinfo}[2]{#2}
\providecommand{\BIBentrySTDinterwordspacing}{\spaceskip=0pt\relax}
\providecommand{\BIBentryALTinterwordstretchfactor}{4}
\providecommand{\BIBentryALTinterwordspacing}{\spaceskip=\fontdimen2\font plus
\BIBentryALTinterwordstretchfactor\fontdimen3\font minus
  \fontdimen4\font\relax}
\providecommand{\BIBforeignlanguage}[2]{{%
\expandafter\ifx\csname l@#1\endcsname\relax
\typeout{** WARNING: IEEEtran.bst: No hyphenation pattern has been}%
\typeout{** loaded for the language `#1'. Using the pattern for}%
\typeout{** the default language instead.}%
\else
\language=\csname l@#1\endcsname
\fi
#2}}
\providecommand{\BIBdecl}{\relax}
\BIBdecl

\bibitem{donahue2018exploring}
C.~Donahue, B.~Li, and R.~Prabhavalkar, ``Exploring speech enhancement with
  generative adversarial networks for robust speech recognition,'' in
  \emph{2018 IEEE International Conference on Acoustics, Speech and Signal
  Processing (ICASSP)}.\hskip 1em plus 0.5em minus 0.4em\relax IEEE, 2018, pp.
  5024--5028.

\bibitem{wang2018supervised}
D.~Wang and J.~Chen, ``Supervised speech separation based on deep learning: An
  overview,'' \emph{IEEE/ACM Transactions on Audio, Speech, and Language
  Processing}, vol.~26, no.~10, pp. 1702--1726, 2018.

\bibitem{pascual2017segan}
S.~Pascual, A.~Bonafonte, and J.~Serra, ``Segan: Speech enhancement generative
  adversarial network,'' \emph{arXiv preprint arXiv:1703.09452}, 2017.

\bibitem{germain2018speech}
F.~G. Germain, Q.~Chen, and V.~Koltun, ``Speech denoising with deep feature
  losses,'' \emph{arXiv preprint arXiv:1806.10522}, 2018.

\bibitem{luo2018tasnet}
Y.~Luo and N.~Mesgarani, ``Tasnet: time-domain audio separation network for
  real-time, single-channel speech separation,'' in \emph{2018 IEEE
  International Conference on Acoustics, Speech and Signal Processing
  (ICASSP)}.\hskip 1em plus 0.5em minus 0.4em\relax IEEE, 2018, pp. 696--700.

\bibitem{rethage2018wavenet}
D.~Rethage, J.~Pons, and X.~Serra, ``A wavenet for speech denoising,'' in
  \emph{2018 IEEE International Conference on Acoustics, Speech and Signal
  Processing (ICASSP)}.\hskip 1em plus 0.5em minus 0.4em\relax IEEE, 2018, pp.
  5069--5073.

\bibitem{stoller2018wave}
D.~Stoller, S.~Ewert, and S.~Dixon, ``Wave-{U-net}: A multi-scale neural
  network for end-to-end audio source separation,'' \emph{arXiv preprint
  arXiv:1806.03185}, 2018.

\bibitem{yao2019coarse}
J.~Yao and A.~Al-Dahle, ``Coarse-to-fine optimization for speech enhancement,''
  \emph{arXiv preprint arXiv:1908.08044}, 2019.

\bibitem{choi2019phase}
H.-S. Choi, J.-H. Kim, J.~Huh, A.~Kim, J.-W. Ha, and K.~Lee, ``Phase-aware
  speech enhancement with deep complex {U}-net,''
  \emph{https://openreview.net/forum?id=SkeRTsAcYm, retracted}, 2019.

\bibitem{wang2014training}
Y.~Wang, A.~Narayanan, and D.~Wang, ``On training targets for supervised speech
  separation,'' \emph{IEEE/ACM transactions on audio, speech, and language
  processing}, vol.~22, no.~12, pp. 1849--1858, 2014.

\bibitem{michelsanti2019training}
D.~Michelsanti, Z.-H. Tan, S.~Sigurdsson, and J.~Jensen, ``On training targets
  and objective functions for deep-learning-based audio-visual speech
  enhancement,'' in \emph{ICASSP 2019-2019 IEEE International Conference on
  Acoustics, Speech and Signal Processing (ICASSP)}.\hskip 1em plus 0.5em minus
  0.4em\relax IEEE, 2019, pp. 8077--8081.

\bibitem{williamson2015complex}
D.~S. Williamson, Y.~Wang, and D.~Wang, ``Complex ratio masking for monaural
  speech separation,'' \emph{IEEE/ACM transactions on audio, speech, and
  language processing}, vol.~24, no.~3, pp. 483--492, 2015.

\bibitem{griffin1984signal}
D.~Griffin and J.~Lim, ``Signal estimation from modified short-time fourier
  transform,'' \emph{IEEE Transactions on Acoustics, Speech, and Signal
  Processing}, vol.~32, no.~2, pp. 236--243, 1984.

\bibitem{afouras2018conversation}
T.~Afouras, J.~S. Chung, and A.~Zisserman, ``The conversation: Deep
  audio-visual speech enhancement,'' \emph{arXiv preprint arXiv:1804.04121},
  2018.

\bibitem{oord2016wavenet}
A.~v.~d. Oord, S.~Dieleman, H.~Zen, K.~Simonyan, O.~Vinyals, A.~Graves,
  N.~Kalchbrenner, A.~Senior, and K.~Kavukcuoglu, ``Wavenet: A generative model
  for raw audio,'' \emph{arXiv preprint arXiv:1609.03499}, 2016.

\bibitem{bai2018empirical}
S.~Bai, J.~Z. Kolter, and V.~Koltun, ``An empirical evaluation of generic
  convolutional and recurrent networks for sequence modeling,'' \emph{arXiv
  preprint arXiv:1803.01271}, 2018.

\bibitem{pandey2019tcnn}
A.~Pandey and D.~Wang, ``Tcnn: Temporal convolutional neural network for
  real-time speech enhancement in the time domain,'' in \emph{ICASSP 2019-2019
  IEEE International Conference on Acoustics, Speech and Signal Processing
  (ICASSP)}.\hskip 1em plus 0.5em minus 0.4em\relax IEEE, 2019, pp. 6875--6879.

\bibitem{wilson2018exploring}
K.~Wilson, M.~Chinen, J.~Thorpe, B.~Patton, J.~Hershey, R.~A. Saurous,
  J.~Skoglund, and R.~F. Lyon, ``Exploring tradeoffs in models for low-latency
  speech enhancement,'' in \emph{2018 16th International Workshop on Acoustic
  Signal Enhancement (IWAENC)}.\hskip 1em plus 0.5em minus 0.4em\relax IEEE,
  2018, pp. 366--370.

\bibitem{ephrat2018looking}
A.~Ephrat, I.~Mosseri, O.~Lang, T.~Dekel, K.~Wilson, A.~Hassidim, W.~T.
  Freeman, and M.~Rubinstein, ``Looking to listen at the cocktail party: A
  speaker-independent audio-visual model for speech separation,'' \emph{arXiv
  preprint arXiv:1804.03619}, 2018.

\bibitem{tan2018convolutional}
K.~Tan and D.~Wang, ``A convolutional recurrent neural network for real-time
  speech enhancement.'' in \emph{Interspeech}, 2018, pp. 3229--3233.

\bibitem{jin2014flattened}
J.~Jin, A.~Dundar, and E.~Culurciello, ``Flattened convolutional neural
  networks for feedforward acceleration,'' \emph{arXiv preprint
  arXiv:1412.5474}, 2014.

\bibitem{iandola2016squeezenet}
F.~N. Iandola, S.~Han, M.~W. Moskewicz, K.~Ashraf, W.~J. Dally, and K.~Keutzer,
  ``Squeezenet: Alexnet-level accuracy with 50x fewer parameters and< 0.5 mb
  model size,'' \emph{arXiv preprint arXiv:1602.07360}, 2016.

\bibitem{wang2017factorized}
M.~Wang, B.~Liu, and H.~Foroosh, ``Factorized convolutional neural networks,''
  in \emph{Proceedings of the IEEE International Conference on Computer Vision
  Workshops}, 2017, pp. 545--553.

\bibitem{howard2017mobilenets}
A.~G. Howard, M.~Zhu, B.~Chen, D.~Kalenichenko, W.~Wang, T.~Weyand,
  M.~Andreetto, and H.~Adam, ``Mobilenets: Efficient convolutional neural
  networks for mobile vision applications,'' \emph{arXiv preprint
  arXiv:1704.04861}, 2017.

\bibitem{sandler2018mobilenetv2}
M.~Sandler, A.~Howard, M.~Zhu, A.~Zhmoginov, and L.-C. Chen, ``Mobilenetv2:
  Inverted residuals and linear bottlenecks,'' in \emph{Proceedings of the IEEE
  Conference on Computer Vision and Pattern Recognition}, 2018, pp. 4510--4520.

\bibitem{ioffe2015batch}
S.~Ioffe and C.~Szegedy, ``Batch normalization: Accelerating deep network
  training by reducing internal covariate shift,'' \emph{arXiv preprint
  arXiv:1502.03167}, 2015.

\bibitem{waibel1989phoneme}
A.~Waibel, T.~Hanazawa, G.~Hinton, K.~Shikano, and K.~J. Lang, ``Phoneme
  recognition using time-delay neural networks,'' \emph{IEEE transactions on
  acoustics, speech, and signal processing}, vol.~37, no.~3, pp. 328--339,
  1989.

\bibitem{kingma2014adam}
D.~P. Kingma and J.~Ba, ``Adam: A method for stochastic optimization,''
  \emph{arXiv preprint arXiv:1412.6980}, 2014.

\bibitem{valentini2016investigating}
C.~Valentini-Botinhao, X.~Wang, S.~Takaki, and J.~Yamagishi, ``Investigating
  {RNN}-based speech enhancement methods for noise-robust text-to-speech.'' in
  \emph{SSW}, 2016, pp. 146--152.

\bibitem{valentini2016speech}
------, ``Speech enhancement for a noise-robust text-to-speech synthesis system
  using deep recurrent neural networks.'' in \emph{Interspeech}, 2016, pp.
  352--356.

\bibitem{paszke2015pytorch}
A.~Paszke, S.~Gross, F.~Massa, A.~Lerer, J.~Bradbury, G.~Chanan, T.~Killeen,
  Z.~Lin, N.~Gimelshein, L.~Antiga, A.~Desmaison, A.~Kopf, E.~Yang, Z.~DeVito,
  M.~Raison, A.~Tejani, S.~Chilamkurthy, B.~Steiner, L.~Fang, J.~Bai, and
  S.~Chintala, ``Pytorch: An imperative style, high-performance deep learning
  library,'' in \emph{Advances in Neural Information Processing Systems 32},
  H.~Wallach, H.~Larochelle, A.~Beygelzimer, F.~d'Alche Buc, E.~Fox, and
  R.~Garnett, Eds.\hskip 1em plus 0.5em minus 0.4em\relax Curran Associates,
  Inc., 2019, pp. 8024--8035.

\bibitem{taal2011algorithm}
C.~H. Taal, R.~C. Hendriks, R.~Heusdens, and J.~Jensen, ``An algorithm for
  intelligibility prediction of time--frequency weighted noisy speech,''
  \emph{IEEE Transactions on Audio, Speech, and Language Processing}, vol.~19,
  no.~7, pp. 2125--2136, 2011.

\bibitem{rec2005p}
ITU, ``P. 862.2: Wideband extension to recommendation p. 862 for the assessment
  of wideband telephone networks and speech codecs,'' \emph{International
  Telecommunication Union, CH--Geneva}, 2005.

\bibitem{he2016deep}
K.~He, X.~Zhang, S.~Ren, and J.~Sun, ``Deep residual learning for image
  recognition,'' in \emph{Proceedings of the IEEE conference on computer vision
  and pattern recognition}, 2016, pp. 770--778.

\bibitem{he2016identity}
------, ``Identity mappings in deep residual networks,'' in \emph{European
  conference on computer vision}.\hskip 1em plus 0.5em minus 0.4em\relax
  Springer, 2016, pp. 630--645.

\bibitem{szegedy2015going}
C.~Szegedy, W.~Liu, Y.~Jia, P.~Sermanet, S.~Reed, D.~Anguelov, D.~Erhan,
  V.~Vanhoucke, and A.~Rabinovich, ``Going deeper with convolutions,'' in
  \emph{Proceedings of the IEEE conference on computer vision and pattern
  recognition}, 2015, pp. 1--9.

\bibitem{szegedy2016rethinking}
C.~Szegedy, V.~Vanhoucke, S.~Ioffe, J.~Shlens, and Z.~Wojna, ``Rethinking the
  inception architecture for computer vision,'' in \emph{Proceedings of the
  IEEE conference on computer vision and pattern recognition}, 2016, pp.
  2818--2826.

\bibitem{szegedy2017inception}
C.~Szegedy, S.~Ioffe, V.~Vanhoucke, and A.~A. Alemi, ``Inception-v4,
  inception-resnet and the impact of residual connections on learning,'' in
  \emph{Thirty-First AAAI Conference on Artificial Intelligence}, 2017.

\bibitem{goodfellow2014generative}
I.~Goodfellow, J.~Pouget-Abadie, M.~Mirza, B.~Xu, D.~Warde-Farley, S.~Ozair,
  A.~Courville, and Y.~Bengio, ``Generative adversarial nets,'' in
  \emph{Advances in neural information processing systems}, 2014, pp.
  2672--2680.

\bibitem{isola2017image}
P.~Isola, J.-Y. Zhu, T.~Zhou, and A.~A. Efros, ``Image-to-image translation
  with conditional adversarial networks,'' in \emph{Proceedings of the IEEE
  conference on computer vision and pattern recognition}, 2017, pp. 1125--1134.

\bibitem{fu2019metricgan}
S.-W. Fu, C.-F. Liao, Y.~Tsao, and S.-D. Lin, ``Metricgan: Generative
  adversarial networks based black-box metric scores optimization for speech
  enhancement,'' \emph{arXiv preprint arXiv:1905.04874}, 2019.

\end{thebibliography}

\end{document}